\documentclass{article}

\usepackage{graphicx} 
\usepackage{subfigure}

\usepackage{natbib}

\usepackage{amsmath,epsfig,amssymb}
\usepackage{hyperref}

\usepackage[accepted]{icml2011}

\icmltitlerunning{Exploiting Correlation in Sparse Signal Recovery Problems}

\begin{document}

\twocolumn[
\icmltitle{Exploiting Correlation in Sparse Signal Recovery Problems: Multiple Measurement Vectors, Block Sparsity, and Time-Varying Sparsity}

\icmlauthor{Zhilin Zhang}{z4zhang@ucsd.edu}
\icmlauthor{Bhaskar D. Rao}{brao@ucsd.edu}
\icmladdress{ECE Department,University of California at San Diego, La Jolla, CA 92093-0407, USA}


\vskip 0.3in
]


\section{Introduction}
\label{sec:introduction}

A trend in compressed sensing (CS) is to exploit structure for improved reconstruction performance. In the basic CS model (i.e. the single measurement vector model), exploiting the clustering structure among nonzero elements in the solution vector has drawn much attention, and many algorithms have been proposed such as group Lasso \cite{groupLasso}. However, few algorithms explicitly consider correlation within a cluster. Meanwhile, in the multiple measurement vector (MMV) model \cite{Cotter2005} correlation among multiple solution vectors is largely ignored. Although several recently developed algorithms consider the exploitation of the correlation, such as the Kalman Filtered Compressed Sensing (KF-CS) \cite{Vaswani08}, these algorithms  need to know a priori the correlation structure, thus limiting their effectiveness in practical problems.

Recently, we developed a sparse Bayesian learning (SBL) algorithm, namely \textbf{T-SBL}, and its variants \cite{Zhilin_2011IEEE,Zhilin_ICASSP2011,Zhilin_ICASSP2010}, which adaptively learn the correlation structure and exploit such correlation information to significantly improve reconstruction performance. Here we establish their connections to other popular algorithms, such as the group Lasso, iterative reweighted $\ell_1$ and $\ell_2$ algorithms, and algorithms for time-varying sparsity. We also provide strategies to improve these existing algorithms.

\section{T-SBL: Exploiting Correlation in the MMV Model}
\label{sec:TSBL}

The MMV model  is expressed as:
\begin{eqnarray}
\mathbf{Y}= \mathbf{\Phi} \mathbf{X} + \mathbf{V}.
\label{equ:MMV basicmodel}
\end{eqnarray}
Here $\mathbf{Y}\triangleq[\mathbf{Y}_{\cdot1},\cdots,\mathbf{Y}_{\cdot L}] \in \mathbb{R}^{N \times L}$ is an available measurement matrix consisting of $L$ measurement vectors. $\mathbf{\Phi} \in \mathbb{R}^{N \times M} (N \ll M)$ is a known dictionary matrix, and any $N$ columns of $\mathbf{\Phi}$ are linearly independent. $\mathbf{X}\triangleq [\mathbf{X}_{\cdot 1},\cdots,\mathbf{X}_{\cdot L}] \in \mathbb{R}^{M \times L}$ is an unknown and full column-rank solution matrix. A key assumption here is that $\mathbf{X}$ has only a few nonzero rows (i.e. the common sparsity assumption \cite{Cotter2005}). $\mathbf{V}$ is an unknown noise matrix.

Most existing algorithms ignore the correlation structure in each row of $\mathbf{X}$. In contrast, T-SBL considers such correlation by assuming the joint density of each row vector of $\mathbf{X}$ to be
\begin{eqnarray}
p(\mathbf{X}_{i\cdot}; \gamma_i,\mathbf{B}_i) \sim \mathcal{N}(\mathbf{0},\gamma_i \mathbf{B}_i), \quad  i=1,\cdots,M
\nonumber
\end{eqnarray}
where $\gamma_i$ is a nonnegative hyperparameter determining whether the $i$-th row $\mathbf{X}_{i\cdot}$ is zero or not. $\mathbf{B}_i$ is a positive definite  matrix that captures the  correlation structure of $\mathbf{X}_{i\cdot}$.

By letting $\mathbf{y}=\mathrm{vec}(\mathbf{Y}^T) \in \mathbb{R}^{NL \times 1}$, $\mathbf{D} = \mathbf{\Phi} \otimes \mathbf{I}_L$, $\mathbf{x}=\mathrm{vec}(\mathbf{X}^T) \in \mathbb{R}^{ML \times 1}$, and $\mathbf{v}=\mathrm{vec}(\mathbf{V}^T)$, we can transform the MMV model (\ref{equ:MMV basicmodel})  to the following one
\begin{eqnarray}
\mathbf{y}= \mathbf{D} \mathbf{x} + \mathbf{v},
\label{equ:blocksparsemodel}
\end{eqnarray}
where $\mathbf{x}$ is block-sparse with each block being $\mathbf{x}_i \in \mathbb{R}^{L \times 1}$, i.e, $\mathbf{x}=[\mathbf{x}_1^T,\cdots,\mathbf{x}_M^T]^T$. Here $\otimes$ indicates the Kronecker product, and $\mathrm{vec}(\cdot)$ is the vectorization operator.

In the SBL framework \cite{Tipping2001}, the T-SBL algorithm was derived as follows \cite{Zhilin_2011IEEE}:
\begin{eqnarray}
\mathbf{x}  &=& (\lambda \mathbf{\Sigma}_0^{-1} + \mathbf{D}^T \mathbf{D})^{-1} \mathbf{D}^T \mathbf{y}  \nonumber\\
\mathbf{\Sigma}_x &=& \mathbf{\Sigma}_0 - \mathbf{\Sigma}_0 \mathbf{D}^T \big(\lambda \mathbf{I} + \mathbf{D} \mathbf{\Sigma}_0 \mathbf{D}^T \big)^{-1} \mathbf{D} \mathbf{\Sigma}_0 \nonumber\\
\gamma_i & = & \frac{1}{L} \mathrm{Tr}\big[\mathbf{B}^{-1}\mathbf{\Sigma}_x^i \big] + \frac{1}{L} \mathrm{Tr}\big[\mathbf{x}_i^T \mathbf{B}^{-1} \mathbf{x}_i \big], \quad \forall i \label{equ:gamma_i}\\
\mathbf{B} & = & \frac{1}{M} \sum_{i=1}^M \frac{\mathbf{\Sigma}_x^i + \mathbf{x}_i (\mathbf{x}_i)^T}{\gamma_i} \nonumber
\end{eqnarray}
where $\mathbf{\Sigma}_x^{i}$ is the $i$-th principal diagonal block of size $L \times L$ in $\mathbf{\Sigma}_x$. $\mathbf{\Sigma}_0$ is a block diagonal matrix with each block given by $\gamma_i \mathbf{B}$. In this algorithm we assume $\mathbf{B}_i = \mathbf{B}$ ($\forall i$) to avoid overfitting. $\lambda$ is the noise variance, which is also estimated in T-SBL; for clarity we omit its learning rule here (and we also omit such learning rules  in  the following algorithms). A simplified version, which has much less computational load, is also derived in  \cite{Zhilin_2011IEEE}.

\begin{figure}[htb]
\begin{minipage}[b]{.48\linewidth}
  \centering
  \centerline{\epsfig{figure=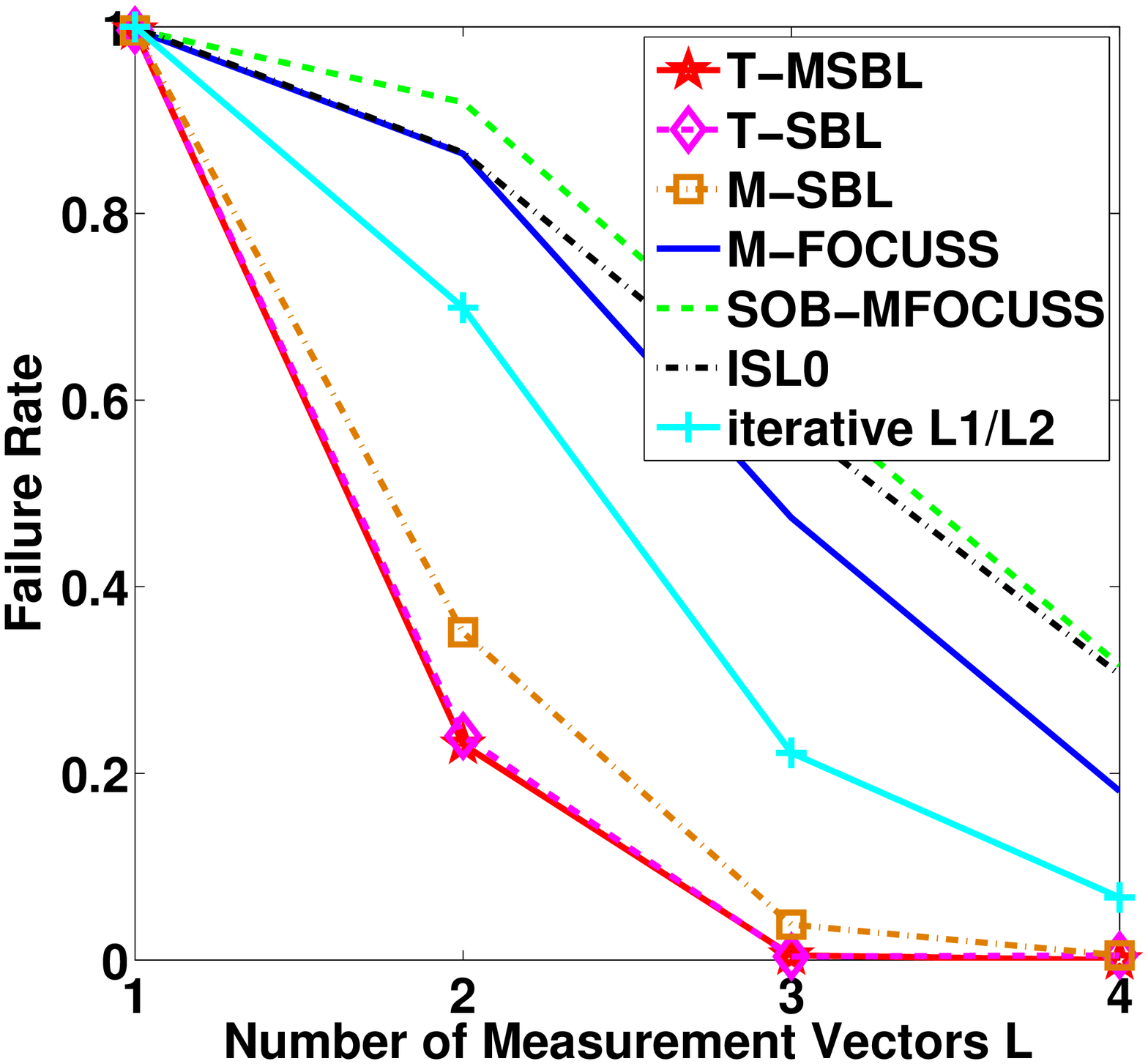,width=4.5cm,height=3.9cm}}
  \centerline{\footnotesize{(a) Correlation: 0}}
\end{minipage}
\hfill
\begin{minipage}[b]{0.48\linewidth}
  \centering
  \centerline{\epsfig{figure=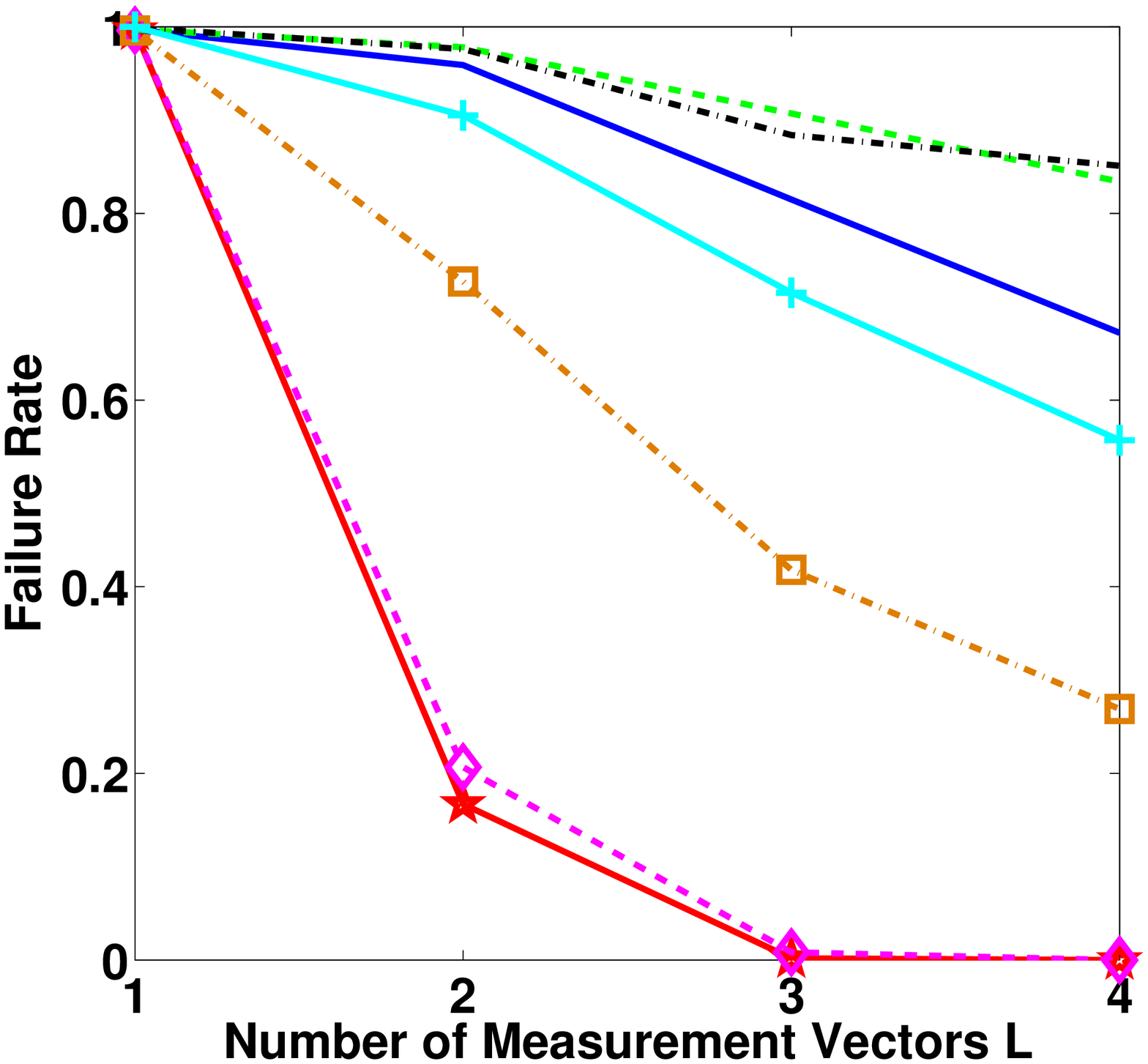,width=4.5cm,height=3.9cm}}
  \centerline{\footnotesize{(b) Correlation: 0.9}}
\end{minipage}
\caption{Failure rates of various algorithms. }
\label{fig:Simulation_varyL}
\end{figure}

We now describe an experiment \cite{Zhilin_2011IEEE} showing that the proposed T-SBL and its simplified version T-MSBL have superior performance when correlation exists among the solution vectors. In the experiment the Gaussian random dictionary matrix $\mathbf{\Phi}$ had the size of $25 \times 125$, the number of nonzero rows of $\mathbf{X}$ was $K=12$, and $L$ varied from 1 to 4. The correlation among solution vectors was 0 and 0.9 in two cases, respectively. Five algorithms were compared (for details see \cite{Zhilin_2011IEEE}) to T-SBL and T-MSBL. To avoid the disturbance of the regularization parameters of all the algorithms, we considered a noiseless case. Results (Fig.\ref{fig:Simulation_varyL}) show that when the solution vectors are highly correlated, all the compared algorithms have very poor performance, due to their inability to exploit such correlation.

In the following we connect T-SBL to other related models.

\section{Connection to Iterative Reweighted $\ell_2$ Framework in the MMV Model}

The iterative reweighted $\ell_2$ minimization framework extended for the MMV problem (in noisy case) computes the solution at the $(k+1)$-th iteration as follows \cite{David2010reweighting}:
\begin{eqnarray}
\mathbf{X}^{(k+1)} = \arg\min_\mathbf{x} \|\mathbf{Y} - \mathbf{\Phi} \mathbf{X}  \|_\mathcal{F}^2 + \lambda \sum_i w_i^{(k)} (\|\mathbf{X}_{i\cdot}\|_q)^2 \label{equ:reweightL2}
\end{eqnarray}
where $w_i^{(k)}$ is the weight depending on the previous estimate of $\mathbf{X}$. Typically $q=2$ or $q=\infty$. In \cite{Zhilin_ICASSP2011} we have shown that T-SBL can be interpreted as an iterative reweighted $\ell_2$ algorithm:
\begin{eqnarray}
\mathbf{X}^{(k+1)} &=& \arg\min_{\mathbf{X}} \Big\{ \|\mathbf{Y}- \mathbf{\Phi} \mathbf{X}  \|_\mathcal{F}^2 + \nonumber \\
&& \lambda \sum_{i=1}^M  (\gamma_i^{(k)})^{-1} \mathbf{X}_{i\cdot} (\mathbf{B}^{(k)})^{-1} \mathbf{X}_{i\cdot}^T \Big\}. \nonumber
\end{eqnarray}
The learning rules for $\gamma_i^{(k)}$ and $\mathbf{B}^{(k)}$ are given in \cite{Zhilin_ICASSP2011}. Note that $\mathbf{X}_{i\cdot} \mathbf{B}^{-1} \mathbf{X}_{i\cdot}^T$ is the quadratic Mahalanobis distance (MD) measure of $\mathbf{X}_{i\cdot}$.

This interpretation reveals the potential advantage of T-SBL is due to using the MD measure of $\mathbf{X}_{i\cdot}$ in  the penalty, instead of using typical $\ell_q$ ($q=2,\infty$) norms of $\mathbf{X}_{i\cdot}$ \cite{Negahban2009}. By comparing it to M-SBL \cite{David2007IEEE}, another SBL algorithm ignoring the correlation in each $\mathbf{X}_{i\cdot}$, we found that T-SBL applies the MD measure also on the weights $(\gamma_i^{(k)})^{-1}$. These observations motivated us to modify existing iterative reweighted $\ell_2$ algorithms for better performance, as shown in \cite{Zhilin_ICASSP2011}.

Although a strict mathematical proof is missing, these empirical results suggest that the mixed norm based penalties as shown in (\ref{equ:reweightL2}) are not very effective for solving the MMV problem in practice, since the unknown
solution vectors are often  correlated.

\section{Connection to Iterative Reweighted $\ell_1$ Framework and Block Sparsity Model}

The iterative reweighted $\ell_1$ minimization framework \cite{Candes2008reweighting} extended for the MMV problem is given by \cite{David2010reweighting}
\begin{eqnarray}
\mathbf{X}^{(k+1)} = \arg\min_\mathbf{x} \|\mathbf{Y} - \mathbf{\Phi} \mathbf{X}  \|_\mathcal{F}^2 + \lambda \sum_i w_i^{(k)} \|\mathbf{X}_{i\cdot}\|_q. \label{equ:reweightL1}
\end{eqnarray}
We now connect T-SBL to this framework.

\begin{figure}
\centering
\includegraphics[width=6cm,height=4.0cm]{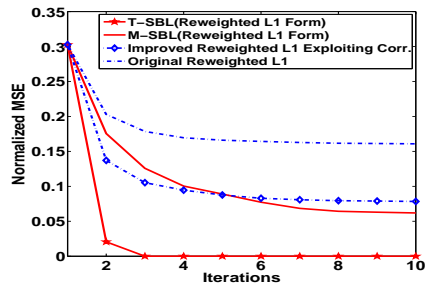}
\caption{Performance improved when exploiting the correlation.}
\label{fig:L1}
\end{figure}

For the model (\ref{equ:blocksparsemodel}) the cost function to estimate all the hyperparameters $\Theta \triangleq \{ \mathbf{B}, \gamma_i,\forall i \}$ is:
\begin{eqnarray}
\mathcal{L}(\Theta) &\triangleq &  -2\log \int p(\mathbf{y}|\mathbf{x};\lambda) p(\mathbf{x};\gamma_i,\mathbf{B}_i,\forall i) d \mathbf{x} \nonumber \\
&=& \log|\lambda \mathbf{I} + \mathbf{D} \mathbf{\Sigma}_0 \mathbf{D}^T  | + \mathbf{y}^T (\lambda \mathbf{I} + \mathbf{D} \mathbf{\Sigma}_0 \mathbf{D}^T)^{-1} \mathbf{y}. \nonumber
\end{eqnarray}
Using the identity $\mathbf{y}^T (\lambda \mathbf{I} + \mathbf{D} \mathbf{\Sigma}_0 \mathbf{D}^T)^{-1} \mathbf{y} \equiv \min_\mathbf{x} \big[\frac{1}{\lambda} \|\mathbf{y}-\mathbf{Dx}\|_2^2 + \mathbf{x}^T \mathbf{\Sigma}_0^{-1} \mathbf{x}  \big]$, we can upper-bound the above cost function as follows:
\begin{eqnarray}
\mathfrak{L}(\mathbf{x},\Theta) = \log|\lambda \mathbf{I} + \mathbf{D} \mathbf{\Sigma}_0 \mathbf{D}^T  | + \frac{1}{\lambda} \|\mathbf{y}-\mathbf{Dx}\|_2^2 + \mathbf{x}^T \mathbf{\Sigma}_0^{-1} \mathbf{x} . \nonumber
\end{eqnarray}
By first minimizing over each member of $\Theta$ and then minimizing over $\mathbf{x}$, we can get the solution:
\begin{eqnarray}
\mathbf{x} = \arg\min_\mathbf{x} \Big\{ \|\mathbf{y}-\mathbf{Dx}\|_2^2 + \lambda g_{\mathrm{TC}}(\mathbf{x}) \Big\},
\label{equ:x_space_expression}
\end{eqnarray}
with  the penalty defined by $g_{\mathrm{TC}}(\mathbf{x}) \triangleq \min_{\mathbf{B} \succ \mathbf{0}, \gamma_i \geq 0,\forall i}  \big\{ \mathbf{x}^T \mathbf{\Sigma}_0^{-1} \mathbf{x}  + \log|\lambda \mathbf{I} + \mathbf{D} \mathbf{\Sigma}_0 \mathbf{D}^T  | \big\}$. Using the duality theory \cite{BoydBook} as in \cite{David2010reweighting}, we can re-express the optimization problem (\ref{equ:x_space_expression}) as follows:
\begin{eqnarray}
\mathbf{x}^{(k+1)} &=& \arg\min_\mathbf{x} \, \|\mathbf{y}-\mathbf{Dx}\|_2^2 \nonumber \\
 && + \lambda \sum_i  w_i^{(k)} \sqrt{ \mathbf{x}_i^T (\mathbf{B}^{(k)})^{-1} \mathbf{x}_i }. \label{equ:RewL1_SMV}
\end{eqnarray}
The learning rules for $w_i^{(k)}$ and $\mathbf{B}^{(k)}$ can be derived using the duality theory and the gradient method. Further, using the approximation in \cite{Zhilin_2011IEEE} we have:
\begin{eqnarray}
\mathbf{X}^{(k+1)} &=& \arg\min_\mathbf{X} \| \mathbf{Y} - \mathbf{\Phi}\mathbf{X} \|_\mathcal{F}^2 \nonumber \\
&& + \lambda \sum_i w_i^{(k)} \sqrt{  \mathbf{X}_{i\cdot} (\mathbf{B}^{(k)})^{-1} \mathbf{X}_{i\cdot}^T  }.
\label{equ:RewL1_MMV}
\end{eqnarray}
The learning rules for  $w_i^{(k)}$ and $\mathbf{B}^{(k)}$ are given by
\begin{eqnarray}
w_i  &\leftarrow& 2 \Big(L \mathbf{\Phi}^T_i \big(\lambda \mathbf{I} + \mathbf{\Phi} \mathbf{\Gamma} \mathbf{\Phi}^T \big)^{-1} \mathbf{\Phi}_i \Big)^{\frac{1}{2}}  \nonumber \\
\mathbf{B} &\leftarrow&  \frac{1}{C} \sum_{i=1}^M  \frac{\mathbf{X}_{i\cdot}^T \mathbf{X}_{i\cdot}}{\gamma_i}, \, \mathrm{with} \,
\gamma_i  \triangleq 2 \frac{\sqrt{\mathbf{X}_{i\cdot} \mathbf{B}^{-1} \mathbf{X}_{i\cdot}^T}}{w_i} \label{equ:learnB}
\end{eqnarray}
where $C\triangleq \sum_{i=1}^M \gamma_i \mathbf{\Phi}_i^T (\lambda \mathbf{I} +\mathbf{\Phi}\mathbf{\Gamma}\mathbf{\Phi}^T )^{-1} \mathbf{\Phi}_i$ and $\mathbf{\Gamma} \triangleq \mathrm{diag}(\gamma_1,\cdots,\gamma_M)$. Note that in each iteration $k$ we need an inner loop to iteratively compute $w_i,\gamma_i$ and $\mathbf{B}$ until convergence for a better estimate of $\mathbf{B}$. The inner loop generally takes several iterations, and the whole algorithm needs very few outer-loop iterations to achieve its best performance (see Fig.\ref{fig:L1}). In fact, each iteration of (\ref{equ:RewL1_MMV}) yields a sparse solution.

When $\mathbf{B}^{(k)} = \mathbf{I}$ and no iteration was performed, the problem (\ref{equ:RewL1_MMV}) reduces to the group Lasso (for the MMV model). When $\mathbf{B}^{(k)} = \mathbf{I}$ ($\forall k$) and iterative reweighting was performed, the problem (\ref{equ:RewL1_MMV}) is a typical iterative reweighted $\ell_1$ algorithm. Thus, T-SBL can be viewed as a variant of iterative reweighted $\ell_1$ algorithms. Similar to the $\ell_2$ interpretation in the previous section, this interpretation also suggests replacing $\ell_q$ norms imposed on $\mathbf{X}_{i\cdot}$ by the MD measure in both the penalty and the weights.

To clearly see the advantage of our suggestion, we conduct the same simulation as in Fig.\ref{fig:Simulation_varyL} (b) when $L=4$. We used the reweighted $\ell_1$ version of T-SBL, the reweighted $\ell_1$ version of M-SBL (which corresponds to the $\ell_1$ version of T-SBL with $\mathbf{B}=\mathbf{I}$) \cite{David2010reweighting}, and the original reweighted $\ell_1$ algorithm (\ref{equ:reweightL1}) with $q=2$ and $w_i^{(k)} = (\| \mathbf{X}_{i\cdot}^{(k)} \|_2+\epsilon)^{-1}$. We also modified this original reweighted $\ell_1$ algorithm to exploit the correlation by changing the weights to :
\begin{eqnarray}
w_i^{(k)} = \Big(\sqrt{\mathbf{X}_{i\cdot}^{(k)} (\mathbf{B}^{(k)})^{-1} (\mathbf{X}_{i\cdot}^{(k)})^T}+\epsilon \Big)^{-1}, \nonumber
\end{eqnarray}
where $\mathbf{B}^{(k)}$ can be estimated by the learning rule (\ref{equ:learnB}). But here we set $\mathbf{B}^{(k)}\;(\forall k)$ to be the true value. The result (Fig.\ref{fig:L1}) shows the algorithms are improved when exploiting the correlation. It is worthwhile to notice that the original iterative reweighted $\ell_1$ algorithm is greatly improved after we replace the $\ell_2$ norm by the MD measure in its weights.

Note that the model (\ref{equ:blocksparsemodel}) is essentially the same as the block sparsity model \cite{groupLasso,Eldar2009} \footnote{Now $\mathbf{D}$ is the original dictionary matrix.}, a variant of the basic CS model. Thus T-SBL can be applied to this model.

\section{Connection to the Time-Varying Sparsity Model}

The time-varying sparsity model is a natural extension of the MMV model. It considers the case when the support of each column of $\mathbf{X}$ is time-varying. Several algorithms have been proposed, such as the Kalman Filtered Based Compressed Sensing (KF-CS) \cite{Vaswani08} and Least-Square Compressed Sensing (LS-CS) \cite{Vaswani10}. Since this model generally assumes the support is changing slowly, we can view such a time-varying  sparsity model as concatenation of several MMV models, where in each MMV model the support does not change. Therefore, T-SBL can be used in this model. Note that in this model exploiting the multiple measurement vectors is important because of the enhanced support-recovery ability afforded by the  MMV model, but unfortunately this strategy is missing in current approaches.

\begin{figure}
\centering
\includegraphics[width=6cm,height=4.1cm]{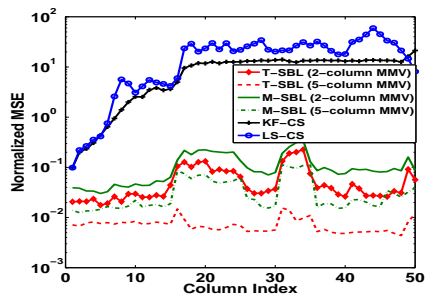}
\caption{Performance in a time-varying sparsity case.}
\label{fig:DCS}
\end{figure}

To verify this strategy, we conduct an experiment using KF-CS, LS-CS, T-SBL and M-SBL. The Gaussian dictionary matrix was of the size $60 \times 256$. The column number of $\mathbf{X}$ was 50. The number of nonzero rows, $K$, during the first 15 columns of $\mathbf{X}$ was 15. $K$ was increased by 10 since the 16-th and the 31-th column of $\mathbf{X}$, respectively. But since the 26-th column 5 nonzero rows were set to zeros. Each nonzero row had temporal correlation varying from 0.7 to 0.99, and had a duration of 20 columns (if was not set to zeros). SNR was about 20 dB. KF-CS and LS-CS were fed with the true noise variance and the true correlation information. However, both T-SBL and M-SBL learned the noise variance. T-SBL also learned the correlation structures. When performing T-SBL and M-SBL, we approximated the time-varying sparsity model using two methods. One was using the concatenation of 25 MMV models, each MMV model containing 2 columns. The second was using 10 MMV models, each containing 5 columns. The experiment was repeated 100 times. Figure \ref{fig:DCS} shows that the two MMV algorithms have better performance than KF-CS and LS-CS. Furthermore, T-SBL is super to M-SBL. The experiment code can be downloaded from the first author's website.

\section{Conclusions}

A general methodology to capture sparsity structure of signals is to use combinations/hierarchy of various norms \cite{Zhao2009}. However, our work showed that another effective way is to use covariance estimation methods to learn the sparsity structures in the framework of SBL. Besides, we showed that iterative reweighted $\ell_1$ and $\ell_2$ algorithms for the MMV model and the block sparsity model can be greatly improved through replacing their $\ell_q$ norms imposed on the blocks/groups by the Mahalanobis distance measure, whose covariance matrix is learned data-adaptively.

\section*{Acknowledgement}
The work was supported by NSF grant CCF-0830612.

\bibliography{sparsebibfile,bookbibfile}
\bibliographystyle{icml2011}

\end{document}